\documentstyle[12pt]{article}
\begin{document}
\addtolength{\baselineskip}{0.5\baselineskip}

\rightline{CALT-68-1784}
\vskip  2cm
\begin{center}
{\large\bf Mass vs. Charge:\\ Quantum Radiation
from Zero Temperature Black Holes\footnote{\rm Work
supported in part by D.O.E Grant
No. DE-FG03-92-ER40701}}
\end{center}
\vskip 1cm
\begin{center}
Jaemo Park and Piljin Yi\footnote{e-mail:
jaemo@cco.caltech.edu and piljin@theory3.caltech.edu}
\end{center}
\begin{center}
 {\em California Institute of Technology, Pasadena, CA 91125}
\end{center}
\vskip 3cm
\centerline{ABSTRACT}
\vskip 1cm
\begin{quote}
We study the mass-charge relation for the semiclassical extremal black hole of
the $S$-wave sector Einstein-Maxwell theory coupled to $N$ conformal scalars.
The classical ratio $M/{|Q|}=1$ is shown to be modified to
$M/{|Q|} \simeq 1-k/6$ for small $ k \equiv  N\hbar/(12\pi Q^{2})$.
Furthermore, numerical study for $k<2$ shows that
$M/|Q|$ is a monotonically decreasing function of $k$.
We speculate on the consequence of such a modification in the 4-dimensional
context.
\end{quote}
\vskip 2cm
\leftline{June, 1993}

\newpage

 In connection with the physics of black hole evaporation, extremal
black holes with vanishing temperature provide interesting theoretical
laboratories. Immune to the Hawking's thermal radiation,
they are the first clues
as to what the final stage of the evaporation process might be.
But this does not mean we can consider the classical extremal black holes
as the final product of the process. For one thing, the thermal behavior
is already expected to break down for near extremal cases\cite{Preskill}.
Zero Hawking temperature simply means the leading quantum effect disappears.
In order to address the questions of black hole quantum physics, we need a
more systematic way of treating quantized matter in nontrivial geometries.
In two-dimensional models, such a method has been adopted by Callan et
al. (CGHS)\cite{Callan}, and used extensively to study 2-D black holes
semiclassically.

In this letter, we want to concentrate on the case of the extremal
Reissner-Nordstr\"{o}m black hole and to study how semiclassical effects
modify one of the classical properties, namely ADM mass $M$. The model we
consider is dimensionally reduced Einstein-Maxwell theory. By restricting
to the spherically symmetric sector we obtain the following 2-D
action,\footnote{$G=c=1$ in this letter}
\begin{equation}
S_{g}= \frac{1}{4}\int d^{2}x\,\sqrt{-g^{(2)}}\, e^{-2\phi}\,
(R^{(2)}+2\,(\nabla \phi)^{2}+2e^{2\phi}-F^{2}),
\end{equation}
where the 4-D metric is split into 2-D metric $g^{(2)}$ and the dilaton part
\begin{equation}
g^{(4)}=g^{(2)}+e^{-2\phi}\,d\Omega^{2}.
\end{equation}
The finite mass solutions with regular horizons are the well-known
Reissner-Nordstr\"{o}m solutions with mass $M$ and charge $Q$ satisfying the
inequality $M \geq |Q|$.

\begin{equation}
g^{(4)}=-F(r)\,dt^{2}+\frac{dr^{2}}{F(r)}+r^{2}\,d\Omega^{2},
\quad F(r)=1-\frac{2M}{r}+\frac{{Q}^{2}}{r^{2}}. \label{eq:RN}
\end{equation}
When the inequality is saturated, $F(r)$ has a double zero at the horizon
$r=M=|Q|$ and the corresponding extremal black hole has zero Hawking
temperature, hence no thermal radiation emanates from the horizon a long
time after the black hole formation. This implies that the usual late time
estimate of the Bogoliubov transformation\cite{Hawking}
is not the leading quantum
correction. It vanishes identically and we need to study next the
nonvanishing
contribution, which may or may not depend on the history of the collapse.

For this purpose, we can follow CGHS and couple $N$ conformal
scalars to the above 2-D action. One can regard these 2-D scalars
as the S-wave part of massless 4-D fermions, alternatively.
Integrating them out completely, which is possible
since we are in a two-dimensional toy world, produces the  non-local
Polyakov-Liouville action\cite{Polyakov} with a particular coefficient,
which summarizes
the effect of the quantized matter on gravity. Furthermore, this
effective semi-classical gravity can be conveniently handled with the
introduction of a  scalar field $z$ with a background charge in the
following manner\cite{Suss},
\begin{equation}
S=S_{g}-\frac{N\hbar}{24\pi} \int d^{2}x\,\sqrt{-g^{(2)}}\,
 ((\nabla z)^{2}-zR^{(2)}).   \label{eq:zaction}
\end{equation}
Since the field equation for $z$ reduces the second term
to the original Polyakov-Liouville action of
central charge $N$, solving this theory at tree level is equivalent to
studying the semi-classical theory of 2-D gravity coupled to $N$ scalars.
These are the leading terms in the large $N$ expansion of the full 2-D quantum
theory, where $N$ is large but $N\hbar$ is of order one.

Notice that the scalar curvature $R^{(2)}$ acts as an external source
coupled to the  $z$ field.
This effectively induces the usual Hawking radiation in a  classical black
hole geometry, i.e.,  a stationary point of $S_{g}$ only.
Because the classical black holes radiate,
the theory does not have any static solution with finite mass and
regular horizon of finite temperature. The only static
solutions of finite mass are those of zero temperature, which were
first studied by S. Trivedi\cite{Trivedi}.

But first let us consider the effect of quantizing the original $N$ conformal
scalars in a classical background. For example, given a classical geometry,
the expectation value of the matter energy momentum tensor can be
found simply by evaluating the classical energy momentum tensor of $z$-field
on that classical background. A family of classical geometries known as
the Vaidya metric\cite{Vaidya} is particulary relevant to our discussion.
\begin{equation}
g^{(4)}=-(1-\frac{2m(v)}{r}+\frac{e^{2}(v)}{r^{2}})dv^{2}+2dv\,dr+
r^{2}\,d\Omega^{2}
\end{equation}
It represents a collapsing massless shell whose cumulative energy and charge
at retarded time $v$ are $m(v)$ and $e(v)$. For smooth $m$ and $e^{2}$,
the cosmic censorship is achieved by requiring the positive
energy condition for the shell\cite{Israel}.\footnote{Of course,
we have introduced an external charged matter source to
create the shell itself.}

For our purposes, however, it is appropriate to choose
\begin{equation}
m(v)=M\theta(v-v_{0}),\quad  e^{2}(v)=Q^{2}\theta(v-v_{0}),
\end{equation}
where $\theta$ is the usual step function. The geometry is then that of an
initial Minkowski spacetime glued to a Reissner-Nordstr\"{o}m black hole
across an
ingoing null shock wave located at $v=v_{0}$. Introducing a new coordinate
$u=v-2 \int F^{-1}(r)\,dr$ with $F(r)$ as in eq.~(\ref{eq:RN}),
$(v,u)$  form a pair of light-cone coordinates above the shock,
\begin{equation}
g^{(2)}=-F(r)\,dvdu,\quad       v>v_{0}.
\end{equation}
In these coordinates, $v\rightarrow \infty$ is the future null infinity, and
$u\rightarrow \infty$ is the future event horizon.
Suppose we impose an initial
condition on the $N$ matter fields such that the expectation value of the
energy-momentum tensor vanishes in the Minkowskian region.
Using energy-momentum tensor conservation,
this can be translated into
\begin{eqnarray}
<T_{uu}>\mid_{v=v_{0}} &=& (\frac{N\hbar}{12\pi}(\partial_{u}^{2}\rho
-(\partial_{u}\rho)^{2})+t_{uu}(u))\mid_{v=v_{0}}=0\\
\rho &=& (\frac{1}{2}\log F), \nonumber
\end{eqnarray}
where $t_{uu}$ comes from the homogeneous part of the solution to $z$ field
equation. This implies the following form of $<T_{uu}>$ as we approach the
future null infinity.
\begin{equation}
<T_{uu}>\mid_{v\rightarrow \infty}
=-t_{uu}(u)=\frac{N\hbar}{12\pi}(\frac{1}{16}F'(r)^{2}-\frac{1}{8}F(r)F''(r))
\mid_{u=v_{0}-2\int F^{-1}\,dr}.
\end{equation}
As $u\rightarrow \infty$, this
clearly  shows a steady flux proportional to the temperature squared
($F'(r\rightarrow \mbox{horizon})\sim T_{Hawking}$).
Also as expected,
this asymptotically steady flux is absent, if $M$ is equal to $|Q|$ so that
$F'(r\rightarrow\mbox{horizon})=0$.
However, there is a finite integrated flux; the total energy radiated is
\begin{equation}
\Delta M=\int_{-\infty}^{\infty}<T_{uu}>\mid_{v\rightarrow \infty}du
=\frac{N\hbar}{96\pi}\int_{|Q|}^{\infty}\frac{F'F'}{F}\,dr
=\frac{k}{6}|Q|,
\end{equation}
where we used $F'(r=|Q|)=0$ for the $M=|Q|$ case. Since we ignored
the gravitational backreaction, this estimate is valid only for small
$k\equiv N\hbar/(12\pi Q^{2})$,
or equivalently for large black holes. Notice that
$\Delta M$ is positive for any $F\geq0$ with a
double zero at the event horizon.
$\Delta M$ represents the energy radiated away by the quantized matter, and
after properly taking into account the gravitational backreaction, the
Bondi Mass of the system should approach as $u\rightarrow \infty$
\begin{equation}
|Q|(1-\frac{k}{6}+O(k^{2})).
\end{equation}
One might assert that it is not clear whether the estimated loss depends on
the particular history of the collapse chosen. After all, the metric chosen
can never be realized, since one cannot assimilate the collapsing
process by a smooth version of the shock wave. As shown in \cite{Israel},
for  smooth $m(v)$ and $e^{2}(v)$ satisfying the positive energy condition,
the extremality can never be achieved in finite time.
It might be that a realistic collapse scenario
produces different $\Delta M$.
We will show that the above estimate of energy loss
is robust by finding numerically the ADM mass of the semi-classical analogue
of the extremal black hole which must be the end
stage of the processs  described so far.

Semi-classical static solutions with
extremal horizons have been studied  near the horizon\cite{Trivedi}.
The requirement of zero temperature horizon specifies a unique initial
condition at the horizon  for a given total charge, and the resulting
static solution is known to be asymptotically flat.
There is no known analytical form of the solution,
but it is, in principle, possible
to carry out numerical integration.

Before going into details of the
simulations performed, it is helpful to discuss other static solutions of
finite mass, all of which have naked singularities.
Those with smaller masses, to be called supercritical,
are qualitatively similar to the classical ones
with $M<Q$. The radius $e^{-\phi}$ monotonically decreases  as we
approach the naked singularity at near origin.
On the other hand, solutions with larger masses, to be called
subcritical, are quite different from classical analogues $M>Q$,
which have curvature singularity at the center $e^{-\phi}=0$
hidden by two layers of nonextremal horizons,
since we assume no heat bath to support nonextremal horizons.
(Heat bath makes ADM mass infinite.) More specifically,
a semi-classical subcritical solution has a lower bound on
the value of the radius $e^{-\phi}$ near would-be horizon.
One can distinguish the two species by observing whether the simulation stops
in the middle or continues all the way to the critical value of the
radius $e^{-\phi_{cr}}\equiv \sqrt{kQ^{2}}$.

 Coming back to the actual simulation, it turns out that static field
equations can be decoupled to produce a single first order differential
equation with the following gauge choice.
\begin{equation}
g^{(2)}=-A^{2}\,dt^{2}+B^{2}Q^{2}\,dr^{2},\quad e^{-2\phi}=Q^{2}r^{2}.
\end{equation}
In this gauge we can extract two independent first order differential
equations.

\begin{eqnarray}
k(\frac{A'}{A})^{2}+2r(\frac{A'}{A})+(1-B^{2}+\frac{B^{2}}{r^{2}})&=&0
\nonumber \\
(r^{2}-k)(\frac{A'}{A}-\frac{B'}{B})+(r^{2}+k)\frac{B^{2}}{r^{3}}-
r(B^{2}-1)&=&0.
\end{eqnarray}
Solving for $A'/A$ in terms of $B$ produces a first order
differential  equation for $1/B^{2}$.
We performed two independent simulations. First, we started
from the asymptotic region with the
initial condition determined by $M/|Q|$, and
searched for the range of $M/|Q|$ producing an extremal horizon (a
double zero of  $1/B^{2}$). Secondly, we integrate outward from the
known behavior  near the extremal horizon and extract the ADM mass
by fitting the  curve in the asymptotic region.
Since the initial points are near, but not quite at the horizon or
$r=\infty$, we needed to calculate accurate initial conditions. Symbolic
expansions of $1/B^{2}$ in appropriate coordinates,
solving the equation above approximately, are
used for this  purpose. Fortunately, the nonanalytic behavior of the metric
near the horizon emphasized in \cite{Trivedi} does not occur for
$1/B^{2}$ as a function of $r$. We used MATHEMATICA for all
numerical and symbolic calculations as well as preparation of the plot.
As we improved the accuracy of the numerical calculation  by
supplying more accurate initial data,  and also by increasing the
intrinsic accuracy of the program used, the results from each
simulation converge to each other. The data for $M/|Q|$ obtained by
the two methods  coincide to an accuracy of $10^{-6}$.

The simulation is carried out only for $k<2$ because the extremal horizon
disappears beyond $k=2$, when the horizon radius is equal to the critical
value of the dilaton $e^{-\phi_{cr}}=\sqrt{kQ^{2}}$.
The plot of $M/|Q|$ as a function of $k\equiv N\hbar/(12\pi Q^{2})$
(Fig. 1) clearly shows the initial slope of $-1/6$ calculated above.
Furthermore, up to $k=2$, the ratio
continues to drop as we increase $N\hbar$ or decrease the charge $|Q|$.

What can we learn from this little demonstration? The first and foremost
fact is that higher order corrections to Hawking's calculation must be taken
account into even for such a crude operation as mass measurement. One should
expect that a similar mechanism works for four-dimensionally black
holes and the
classical bound $M\geq |Q|$ is modified, unless some unbroken extended
supersymmetry protects it.
But the model we used gives few clues as to what the  modification might be.
While 2-D conformal scalars can be interpreted as the $S$-wave modes of
4-D massless fermions, we cannot regard our model as a quantitative
approximation to the full 4-D physics. There is no generic mass gap present
to separate $S$-wave fermions out from the rest.

Nevertheless this doesn't prevent us from speculating on the effect of such
a modified mass-charge relation in 4-D. In particular, suppose the same
monotonic decreasing behavior is realized for the four-dimensionally extremal
black holes. The possibility has been contemplated by J. Preskill
for electrically charged extremal black holes with
emphasis on charge renormalization\cite{Preskill2}.
The most immediate consequence would be to lift the well-known degeneracy
for multi-extremal black hole configurations. Classically, a family
of solutions known as the Papapetrou-Majumdar space-time\cite{pa-maj},
describes many  extremal black holes at rest relative to one another.
The total ADM mass of  such a solution is the sum of the individual masses,
\begin{equation}
M=\sum_{i} |Q_{i}|. \label{eq:mass}
\end{equation}
This can easily be seen by imagining each hole separated from one another
far away, so that whatever potential energy there might be
becomes negligible. In fact, there is no potential between individual
black holes and the total mass is given by eq.~(\ref{eq:mass}) for any finite
separations.
Therefore two different multi-black hole configurations in equilibrium have
the  same energy provided that the sum of absolute value of
the charges are equal. But, with the modified $M/|Q|$
which decreases as $Q^{2}$ decreases,
the same reasoning shows that it is energetically favorable
to split one big black hole into many smaller ones.
The classical degeneracy is lifted.

Classical physics forbids such a bifurcation process, since it violates
the second law of black hole thermodynamics. However, there has been
suggestions of possible finite action instantons mediating bifurcation
of the extremal Reissner-Nordstr\"{o}m black holes. In fact, D. Brill
found an instanton of finite action interpolating between two
Bertotti-Robinson metrics with different numbers of necks\cite{Brill}.
It is  well
known that a Bertotti-Robinson metric with a single neck approximates
an extremal Reissner-Nordstr\"{o}m  black hole near the  horizon.
If the initial and the final states are of the same energy, the instanton
will take infinite Euclidean time to make the transition, and the stationary
state would be a linear combination of the two classical configurations.
With the modified $M/|Q|$ relation however, a relevant Euclidean solution
is a bounce solution and a big extremal Reissner-Nordstr\"{o}m black
would decay to many extremal black holes of smaller charges
distantly separated.\footnote{A similar observation
has been made in the context
of classical dilatonic black holes with a massive dilaton\cite{Horo}.}

So far, we have completely ignored the possible presence of charged
matter fields. Suppose there is an elementary charged particle of
mass $m$ and charge $e$ and consider an extremal black hole of
mass $M$ and charge $Q$. For $m<<|e|$, the Schwinger pair production
near the horizon is always dominant over a possible bifurcation  process
and  the black hole charge will eventually be wiped out.
But for sufficiently large $m>m_{min}$,
it will be kinematically impossible for an extremal black hole to
lose its charge by emiting these charged particles\cite{Preskill2}.
For a large black hole $M\gg m$ in particular, we have
$m_{min}/|e|\simeq M/|Q|$.
Therefore the model we considered should be regarded as a possible
scenario for magnetically charged extremal black holes in a world
where the magnetic monopole comes with mass comparable to, or even
larger than, its charge in Planck units.

In summary we have found that the classical inequality $M\geq |Q|$ can
be modified through semi-classical effects.
It would be most interesting to find out about similar effects
in the context of 4-dimensional models
but it is beyond the scope of this letter.

We would like to thank J. Preskill for critical reading of the manuscript.
Also P.Y. thanks S.Trivedi for invaluable discussions at various stages
of this work.

\newpage

\noindent
{\bf Figure Caption}

\noindent
Fig.1: Plot of $M/|Q|$ versus $k\equiv N\hbar/(12\pi Q^2)$. The straight line
shows the leading behaviour $M/|Q|=1-k/6$. The dots are the actual numerical
results from the two independent simulations. Data points are at
$k=n/10$ for $n=1,...,19$ as well as $k=0.001$.

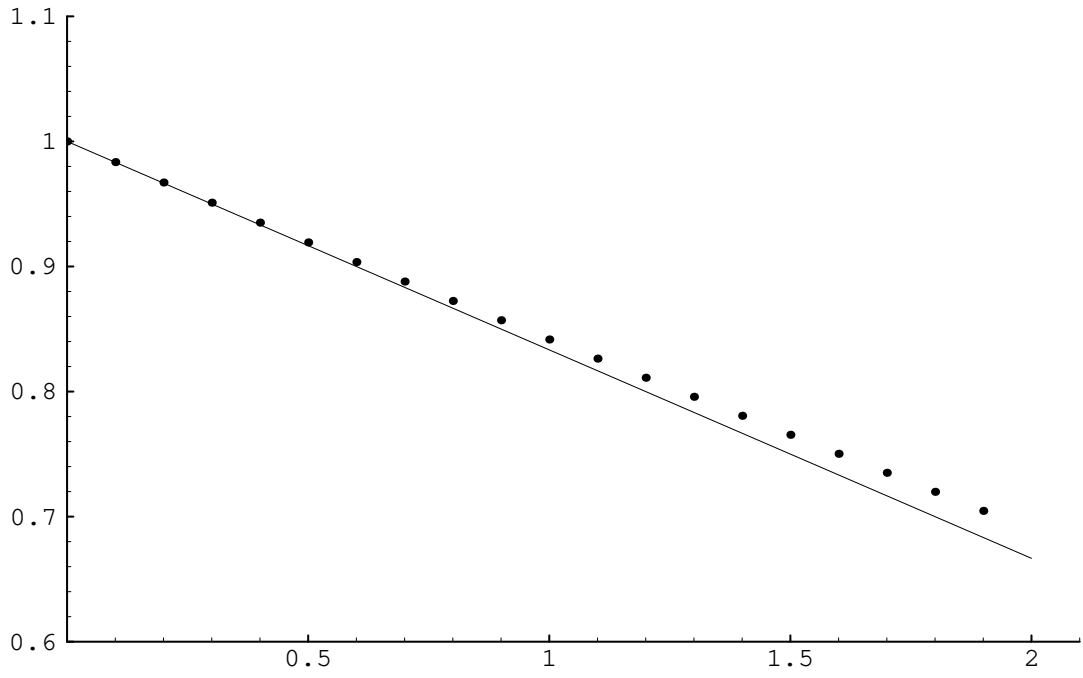
\end{document}